# High Pressure Sputtering For Kigh-*k* Dielectric Deposition. Is It Worth Trying?

E. San Andrés, P. C. Feijoo, M. A. Pampillón, M. L. Lucía, A. del Prado.

Departamento de Física Aplicada III (Electricidad y Electrónica). Facultad de CC. Físicas, Universidad Complutense de Madrid, Avda. Complutense S/N. Madrid, 28040 Spain.

**Abstract**

Our research group studies the deposition of high permittivity dielectrics by a non-standard method: high-pressure sputtering. The dielectrics studied here are gadolinium scandate deposited from dielectric targets, and gadolinium oxide deposited from a metallic target, with an *in situ* plasma oxidation. The stoichiometric gadolinium scandate presents a slight permittivity boost after annealing, but with gadolinium-rich scandate (grown with an annealing of a nanolaminate) the dielectric shows a high effective permittivity of 21 and no noticeable $SiO_2$ layer at the interface with Si. On the other hand, the stacks fabricated with the metallic Gd target have a $SiO_2$ interface less than 0.7 nm thick that can be further reduced by scavenging with Ti gates. In fact, this scavenging effect is also demonstrated for the first time with indium phosphide substrates, obtaining a low capacitance equivalent thickness of only 2.1 nm.

**Introduction**

During the last decade the microelectronics industry has been introducing an increasing amount of changes to the CMOS transistor architecture. One of the biggest changes was the substitution of the gate stack, from the poly Si/SiON structure that was the workhorse until 2007, to the current metal gate/high-*k* stack (1). In this structure, the metal gate is comprised of several layers needed to modulate the effective workfunction, to minimize series resistance, to improve adhesion, etc. The structure of the insulator is also complex, with a sub-nm $SiO_2$ layer under a hafnium oxide based high-*k* dielectric (1). The thickness of these films has to be controlled in the sub-nm range. To achieve this degree of control, the industry has converged on Atomic Layer Deposition (ALD) as the technique of choice for growing the high-*k* dielectric, or even the whole gate stack.

There are many reasons for the predominance of ALD, such as sub-monolayer control of thickness, conformal deposition, uniformity, reproducibility, etc.(2) However, there exist also disadvantages. First, the need of gaseous precursors can cause film contamination, typically from Cl, C and H, since the precursors typically are either chlorides or organometallics. The processing is performed at a moderate temperature (typically 300-500ºC), so completely avoiding the incorporation of these contaminants is difficult. Also, the surface has to be saturated of precursors after each cycle. Thus a very high gas flow is needed to achieve a moderate growth rate. This gas flow prevents working in the high-medium vacuum range and consumes a high amount of precursor. This last fact increases processing costs and, depending on the particular precursor, might also be an environmental hazard. Moreover, the chemisorption of the precursors is critical for layer growth, thus many semiconducting surfaces have to be treated before ALD. In Si a thin chemical oxide is typically used (3). The sub-nm control of thickness comes together with a very low deposition rate, which might limit throughput. Finally, the possibility of obtaining a particular material depends on the availability of precursors, and yet there are not recipes for every material of interest.

Due to these limitations, we firmly believe that there is room for other techniques for high-*k* deposition, which can circumvent some of the ALD drawbacks. In particular, one of these techniques is sputtering, a technology that was used from the very first years of microelectronics for aluminum metallization. At present it is still used in some processes, like copper seed deposition when fabricating the interconnections (4). Sputtering presents some advantages over ALD. For instance, since the target is made of pure material and the sputtering gas is typically inert, film contamination can be minimal.

Complex materials can be deposited as long as a composite target can be fabricated, since the sputtering yields of most materials are in the same order of magnitude. Other approaches used to fabricate compounds are co-sputtering (5) and reactive sputtering (4). Finally, the fabrication of nanolaminates is also possible by sequential deposition (6). Concerning the growth rate, it is fairly constant, and since the chemisorption on the surface is not needed, it is possible to deposit on most starting surfaces. Besides, the target material use efficiency is much better. From an economic point of view this can be relevant for expensive materials. In fact, since the targets suffer from non-uniform sputtering, they must be discarded with a large fraction of the material still remaining. However this material can be recycled after target replacement. Recycling of unused gaseous precursors is much more difficult. Another advantage is that reverse sputtering can be used *in situ* for wafer cleaning. Finally, the deposition temperature can be lower than in ALD (7), so the undesirable $SiO_x$ regrowth under the high-*k* dielectric can be minimized.

However, there are some concerns with sputtering: the main one is that this technique generates the plasma very close to the sample wafer. Electrons and sputtered species present in this plasma can produce degradation of the sample surface by energetic bombardment. Conventional sputtering works in the $10^{-2}$ - $10^{-3}$ mbar range, and the mean free path of the species is in the order of some centimeters (8). On the other hand, High Pressure Sputtering (HPS) works at a higher pressure, typically between 0.5-5 mbar, thus the mean free path is in the order of $10^{-2}$ cm. This way, the sputtered species suffer many collisions before reaching the target, losing their excess energy and becoming thermalized. Once thermalized, they diffuse from the neighborhood of the target and reach the substrate with low energy. Therefore HPS reduces plasma damage while keeping a very compact design (in other words, a low volume vacuum chamber is possible). This is quite important for high wafer throughput. Finally, due to the high pressure that produces many collisions of the sputtered species, the conformality of the films is expected to be better than conventional sputtering. This is key to its applicability to emerging 3D transistor structures.

Concerning the $SiO_2$-like interface, sputtering does not need to grow a thin chemical $SiO_2$ for film nucleation as ALD does (9). However, some substrate oxidation might occur during the first stages of deposition because the substrate is directly exposed to the plasma. Then it is important to develop strategies to minimize this regrowth. A possible solution is to remove the $SiO_x$ interface after deposition using the scavenging effect (10).

HPS was initially developed by Dr. Poppe's group for the growth of epitaxial high temperature superconductors with success (11). Our group adapted the system to deposit many types of high-*k* materials. In this article some recent results on gadolinium and/or scandium based dielectrics will be discussed. There are several reasons to study this family of materials. First of all, $Gd_xSc_{2-x}O_3$, has a crystallization temperature of ~1000ºC (12), and a permittivity in the 20-30 range (13). $Gd_2O_3$ is another high-*k* material of interest: albeit its moderate permittivity, around 15 (14), there are references that show that it can yield functional MOS structures on compound semiconductors (15). The importance of this lies in the current effort to develop device quality high-*k* dielectrics on III-V semiconductors (for n-MOS fabrication) (16) and on Ge or $Si_xGe_{1-x}$ (for p-MOS devices) (17).

We will show results of $Gd_xSc_{2-x}O_3$ deposited from a single $GdScO_3$ compound target, and will be compared with the data obtained from a sequential deposition of $Gd_2O_3$ and $Sc_2O_3$. Also we will discuss results obtained with $Gd_2O_3$ but using a metallic target and a two-step deposition procedure with the aim to minimize $SiO_x$ regrowth. The characterization of this process on Si and InP high mobility wafers will be shown.

**Experimental**

The HPS system that was used to deposit the high-*k* dielectrics consists of a small chamber made from standard high-vacuum pieces. This chamber reaches an ultimate pressure below $10^{-6}$ mbar. It is equipped with a robotic arm that can hold up to three different targets, each attached to an independent *rf* source. The targets have a diameter of 4.5 cm and are 0.3 cm thick. The purity is the maximum available, 99.95%, and the targets used were stoichiometric $GdScO_3$ for single target deposition, $Gd_2O_3$ and $Sc_2O_3$ for nanolaminated $Gd_xSc_{2-x}O_3$, and metallic Gd for interface minimization. The robotic arm is controlled by a computer, allowing the fabrication of laminates by sequential sputtering. The sample holder and the targets are located in a parallel-plate configuration, with a target-to-substrate distance of 2.5 cm. The maximum wafer diameter is 2", and the substrate holder temperature can be controlled from room temperature to 700ºC, but all depositions were performed either at 200ºC (when depositing $Gd_xSc_{2-x}O_3$) or at room temperature ($GdO_x$ growth). Vacuum is obtained by a turbomolecular pump. During depositions the pumping speed is limited by a guillotine valve to avoid the need of very high gas flows and pump overheating. The depositions were performed at pressures between 0.5 and 1.5 mbar. The system has a gas delivery system regulated by mass flow controllers with high purity Ar and $O_2$ lines. The plasma was excited by a Hüttinger *rf* power supply working at a frequency of 13.56 MHz attached to an impedance matchbox. The incident powers used in these studies were between 10 and 60 W.

In the first set of samples that will be presented, the stoichiometric $GdScO_3$ target was used. In all experiments with oxide targets the sputtering atmosphere used was pure Ar, since previous results showed that the oxygen from the target was enough to produce stoichiometric films (18,19). The substrate holder temperature was 200ºC, the chamber pressure was 1.0 mbar and the *rf* power 40 W.

To improve the properties of these scandates, nano-laminates of $Gd_2O_3$ and $Sc_2O_3$ were fabricated, with the aim to obtain $Gd_xSc_{2-x}O_3$ after annealing. The composition was controlled by adjusting the relative thicknesses of $Gd_2O_3$ and $Sc_2O_3$. In the design of these experiments we used the pressure and *rf* power that produced the best results according to previous works on $Sc_2O_3$ and $Gd_2O_3$ (20,21). Thus, the laminates were grown at an Ar pressure of 1.0 mbar and *rf* power of 40 W, with a substrate temperature of 200ºC. We aimed to Gd-rich scandate for two reasons: firstly, $Gd_2O_3$ presented a thinner $SiO_x$ interface than $Sc_2O_3$ and thus it possesses a higher stability with Si; secondly, according to Kittl *et al.* (22), Gd-rich $Gd_xSc_{2-x}O_3$ has a higher relative permittivity than other compositions (with a value slightly above 25). During the whole deposition process power was applied to both targets simultaneously using a separate *rf* source for each cathode.

For interface regrowth control, we have studied a different approach. Instead of using an oxide target or a metal target and a reactive sputtering, we have developed a two-step procedure: first we cover the substrate with a thin metallic Gd film by sputtering Gd in pure Ar, and afterwards we perform a plasma oxidation in an $Ar/O_2$ atmosphere (95%/5%). The introduction of oxygen to the gas mixture changes abruptly the plasma dynamics, inhibiting Gd sputtering, and the effect of this step is the plasma oxidation of the metallic Gd. With this procedure we avoid exposing the semiconductor to oxygen, but it is critical to adjust the Gd thickness to the plasma oxidation time and *rf* power. If the oxidation is too mild the result can be a defective $GdO_x$ film. On the other hand, an aggressive plasma oxidation might oxidize not only the Gd film but also the underlying substrate. These processes were carried out at room temperature and the pressure was 0.5 mbar. The *rf* power used for metal deposition was 30 W. The Gd film deposition time was varied from 80 s to 300 s, and the oxidation *rf* power and duration was adjusted between 10 – 30 W and 100 - 600 s respectively.

For physical characterization of the high-*k* films we used high resistivity *n*-Si (100) 2" wafers, with a resistivity of 200 – 1000 Ω cm and polished on both sides. On the other hand, for electrical characterization of the films, metal-insulator-semiconductor (MIS) devices were fabricated on single side polished (100) *n*-Si wafers with lower resistivity of 1.5 - 5.0 Ω cm. In order to be applicable to the 7 nm node and beyond, HPS must be able to deposit the dielectrics on high mobility substrates. To demonstrate this feasibility we grew GdO$_x$ on undoped InP substrates (thus *n*-type behavior) with (100) orientation and a resistivity of 1.0 – 5.0 Ω cm. On the substrates used for MIS fabrication, square openings were defined on field oxide. The Si samples were cleaned by a standard RCA (*Radio Corporation of America*) process, while the InP substrates were prepared with a 1 min immersion in iodic acid diluted to 10% by weight (23). Just before the introduction to the HPS chamber both substrates were etched in a 1:50 HF solution for 30 s to remove the native oxides. After high-*k* deposition the gate electrode was e-beam evaporated and defined by a lift-off procedure. Several metal stacks were studied, composed of Pt, Ti and Al with different thicknesses. Backside contact was also e-beam evaporated (Ti/Al for Si devices, AuGe/Au for InP).

Several physical characterization techniques were used: the bonding structure of the dielectric films was analyzed by Fourier transform infrared (FTIR) spectroscopy using a Nicolet Mangna-IR 750 Series II in transmission mode at normal incidence. The dielectric film absorbance spectrum was obtained after subtraction a bare-Si substrate spectrum; X-ray photoelectron spectroscopy (XPS) spectra were obtained by a VG Escalab 200R spectrometer equipped with a MgKα X-ray source (hν=1254.6 eV), powered at 120W; transmission electron microscopy (TEM) lamellas were prepared by a DualBeam focused ion beam system from FEI, while high resolution TEM images were taken with a Tecnai T20 microscope also from FEI at an energy of 200 keV. The TEM microscope is equipped with an X-ray detector that obtains the energy-dispersive X ray (EDX) spectra when the system works in the scanning-TEM configuration.

Concerning electrical characterization, we measured the high frequency $C_{HF}$-$V_G$ and $G$-$V_G$ curves of the MIS devices using an Agilent 4294A impedance analyzer and the leakage current density ($J_G$-$V_G$) using a Keithley 2636A system. Most samples were measured before and after forming gas annealings.

Unless specified, these annealings were performed for 20 min at temperatures between 300ºC and 500ºC. We extracted parameters such as the capacitance equivalent thickness (CET), the equivalent oxide thickness (EOT), the flatband voltage ($V_{FB}$) or the interface trap density ($D_{it}$). This way, we analyzed the influence of the FGA on the dielectric permittivity and the electrical performance.

**Results and discussion**

Fig. 1 shows the FTIR spectrum of $Gd_xSc_{2-x}O_3$ films deposited on Si from a single stoichiometric $Gd_xSc_{2-x}O_3$ target. The samples were sputtered during variable times, between 10 min and 60 min, keeping the same deposition conditions. Although the spectra were measured up to a wavenumber of 4000 cm$^{-1}$, no relevant features were found for wavenumbers above 1200 cm$^{-1}$. Besides the C-O absorption at 667 cm$^{-1}$, that is related to the $CO_2$ variations in the measurement atmosphere, and some substrate correction artifacts in the 400-650 cm$^{-1}$ range, there are two features that can be clearly related to the film. The clearer is a broad peak in the 400-550 cm$^{-1}$ range. The area of this peak is roughly proportional to the deposition time, so we can conclude that this feature is due to the $Gd_xSc_{2-x}O_3$ film. In fact, for the thicker films, two peaks are clearly observed at the same vibration wavenumbers previously found in $Gd_2O_3$ and $Sc_2O_3$ (24,25), as highlighted in the figure. We can conclude thus that HPS is able to extract Gd and Sc to produce $Gd_xSc_{2-x}O_3$ with a constant growth rate. On the other hand, there is another much less intense feature with its maximum at 1025 cm$^{-1}$, as can be seen in the inset of fig.1. The shape and intensity of this peak is independent on the sputtering duration, and its maximum is close to the wavenumber of the relaxed O-Si-O stretching vibration, 1065 cm$^{-1}$ (26). This feature indicates that there is some $SiO_x$ formation at the Si/high-*k* interface, most likely strained $SiO_2$. Since the peak is almost identical for all films then either it grows during the first stages of the deposition, while Si is exposed to the oxygen containing plasma, or it is due to the reaction of $Gd_xSc_{2-x}O_3$ with Si. We performed annealings at temperatures up to 450ºC and we did not observe any change on the spectra. This fact points to the plasma oxidation theory as the reason for the SiO$_x$ regrowth.

Focusing on the thinner films, we fabricated MIS devices on Si with thin Gd$_x$Sc$_{2-x}$O$_3$ deposited with the same plasma conditions but sputtered during 5, 10 and 15 min. The gate was pure Pt (25 nm) to avoid the reaction of the gate electrode with the high-*k*. In fig. 2 we show the small signal capacitance as a function of gate voltage of the thinnest sample. The curves were measured before any thermal treatment, and after annealing at 300ºC and 450ºC. We wanted to observe if there is a permittivity enhancement after annealing or an interfacial regrowth. We found that in all three thicknesses (not shown) there is a slight increase in capacitance when annealing at 300ºC, with some variability between devices. On the other hand, at 450ºC the increase is clearer in all devices, indicating a moderate permittivity boost. By fitting the EOT of the devices as a function of film thickness we obtain that at 450ºC the permittivity is about 15 with an interfacial layer 1.5 nm thick. However, since the physical thickness was interpolated from the ellypsometric measurements of thicker films and we do not have TEM measurements of these samples, these values have to be taken with caution. In any case, these results suggest that depositing from a single stoichiometric target has limitations both in permittivity and interfacial regrowth.

To overcome these limitations, we deposited Gd$_x$Sc$_{2-x}$O$_3$ nano-laminates aiming at Gd-rich compositions, which the literature (22) shows that present the highest permittivity. Also we had found that HPS deposited Gd$_2$O$_3$ films presented a thinner interface in contact with Si than Sc$_2$O$_3$ (27). Thus, the first layer of the laminate was always Gd$_2$O$_3$, aiming at a total thickness of the stack of 6 nm with 8 Gd$_2$O$_3$ /Sc$_2$O$_3$ deposition cycles.

The XPS results of this process are shown in fig. 3, where it is clear that the sample contains both Sc and Gd. By fitting the XPS peaks and correcting by the sensitivity factors (1.6 for Sc 2p and 3.41 for Gd 3d$_{5/2}$) we can calculate composition. The result is that the scandate has a Gd$_{1.8}$Sc$_{0.2}$O$_3$ composition, so it is a Gd-rich scandate as intended. However, these measurements give information on the composition of the top region of the film, but do not give insight on the nanostructure of the film, to discern if it is a laminate or if there is intermixing of the binary dielectrics.

To gain insight on the structure of the films we performed high resolution TEM measurements. The film cross-section image after an annealing at 450ºC in forming gas is shown in fig. 4. This picture shows an amorphous dielectric with a thickness of 7±1 nm, sandwiched between mono-crystalline Si and a poly-crystalline Pt layer. The Pt/ $Gd_{1.8}Sc_{0.2}O_3$ interface is quite abrupt, as expected, since Pt is a noble metal. No clear interfacial $SiO_x$ layer between the $Gd_{1.8}Sc_{0.2}O_3$ and the Si substrate is observable. On the other hand, no multi-layer structure was observed (the dielectric was an 8-cycle film), but there are two different intensities within the dielectric layer. This fact, together with EDX results, suggest a reaction between the dielectric layer and the thin interface $SiO_x$, possibly developing a gadolinium-scandium silicate with a thickness of around 3.8 nm. This reaction may take place during the annealings and could limit the effective permittivity of the dielectric, since silicates generally present lower dielectric constants than the pure oxides.

The aim of the annealing was to intermix the $Gd_2O_3$ and $Sc_2O_3$ nano-layers in order to develop the high-*k* phase. Fig. 5 shows the gate capacitance as a function of gate voltage before and after the sequential annealings of the same wafer. We found a significant accumulation capacitance boost, by a factor of more than 2. This means that the dielectric is stable in contact with Si, since if there were $SiO_x$ regrowth the capacitance would decrease. The most striking result is that the effective permittivity of the dielectric is increasing, which suggests that annealing promotes the transition from nano-laminate to $Gd_{1.8}Sc_{0.2}O_3$, decreasing the EOT from 3.0 nm to 1.3 nm. The calculated effective relative permittivity of the dielectric layer is ~21, using the measured thickness of 7 nm, indicating that the silicate has only a moderate impact on permittivity.

As explained in the experimental section, in parallel with this work we have used a two-step approach to control the interface regrowth: metallic Gd deposition followed by *in situ* plasma oxidation. We have optimized this process on Si, and some electrical results are shown in Fig. 6. In this sample the metal was deposited for 80 s and followed by a short plasma oxidation of only 100 s. Ellipsometric measurements showed that this process produced a 6.0 nm thick $Gd_2O_3$ film. In figure 6 we show that there is a slight increase in the accumulation capacitance values before and after FGA up to 400ºC, indicating a stable $Gd_2O_3$ on Si (which means some permittivity increase). The EOT of these samples after annealing at 400ºC is 2.2 nm. In the extreme case that the plasma oxidized $Gd_2O_3$ permittivity would be 15 [28], then the interfacial thickness would be of only 0.68 nm. This sub-nm interface is very promising, since it comes with a stable high-*k* and a low density of interface states, as can be appreciated from the decrease of the hump in depletion and the steeper curve after the annealing.

To further decrease the interfacial oxide, we studied the possibility of using the interface scavenging effect (29) with our HPS deposited samples. Fig. 7 shows the HRTEM images of a non-optimized $Gd_2O_3$ sample that was obtained after 80 s metal deposition and 300 s plasma oxidation. Half of the wafer was covered with Pt, and the other half with thick Ti, and annealed at 300ºC afterwards. We can observe that the Ptcapped half presents a 1.5±0.5 nm $SiO_2$-like interface under the 6.0±1.0 nm $Gd_2O_3$ film. On the other hand, the Ti-capped half shows that the $SiO_2$ layer has disappeared, indicating that indeed Ti is scavenging the interface. However, there is also a marked reduction in $Gd_2O_3$ thickness, suggesting that when Ti runs out of $SiO_2$, it commences to consume also $Gd_2O_3$. Thus, we have proven that high pressure sputtered $Gd_2O_3$ is compatible with the interface scavenging concept. Nevertheless, to effectively use this effect it is critical to optimize both scavenging layer thickness and the annealing processes.

Going one step further, we studied the compatibility of the plasma oxidation process with more delicate III-V semiconductors. In the HRTEM image of fig. 8 we present the cross section of a $Gd_2O_3$ film deposited on InP and capped with a 4 nm Pt / 25 nm Al electrode, after a FGA at 400ºC for 20 min. The dielectric film was sputtered for 120 s and plasma oxidized for 200 s so the thickness should be ~9 nm. In the image we can observe that there are no clear intermediate oxides at the $Gd_2O_3$/InP interface. This is a very interesting result, since this process applied to Si produces a thin $SiO_x$ layer at the interface. This fact points to a superior oxidation resistance of InP when exposed to the $Ar/O_2$ plasma and/or in contact with $Gd_2O_3$. The InP surface seems quite rough. It cannot be discerned whether this is an image effect due to the absence of interfacial oxides together with the polycrystalline character of the $Gd_2O_3$, a reaction of the dielectric with InP, or just a rough InP starting surface. Here the dielectric presents a polycrystalline character with a thickness of ~8 nm, as opposed to the amorphous structure when the deposition is performed on Si. On top of this layer we observe a brighter layer ~4 nm thick. To clarify the origin of this layer we performed EDS analysis of the image (30), and we found that there was an intermixing of the Pt/Al electrode together with Al diffusion to the $Pt/Gd_2O_3$ interface, producing the bright GdAlO layer. In any case, assuming a CET of the semiconductor accumulation layer of 0.7 nm and no interface between the dielectric and InP, the effective permittivity of the stack is 13±1, which is very close to the stoichiometric value. Thus the effect of the aluminate layer is not critical. Besides, this value confirms a fully oxidized $Gd_2O_3$.

Finally, fig. 9 shows the capacitance-voltage curves of metal-insulator-semiconductor $Gd_2O_3$/InP devices with three different gate stacks. In this set of curves we compare the results of the Pt/Al MIS devices depicted of figure 8 with the results obtained by adjusting several processing parameters: reducing the plasma oxidation time to 100 s, avoiding Al capping for Pt samples and reducing FGA temperature to 325ºC. Also, to explore the scavenging effect on InP we fabricated a sample with a metal electrode stack composed of 5 nm Ti capped with 25 nm of Pt.

On all samples we found a full accumulation-depletion-inversion sweep, which is an indication of an unpinned Fermi level. It is noteworthy that, to our knowledge, only T. Das *et al*.(31) have published results of functional MIS devices deposited on III-V by sputtering. This result shows that HPS together with the optimized plasma oxidation is a process that preserves the delicate III-V semiconductor surface.

Finally, the $C_{HF}$-$V_G$ curves for the Ti/Pt electrode sample show a capacitance increase of around 15%. This result demonstrates that the scavenging effect can also be applied to III-V substrates. This is very relevant from an integration standpoint, since in the near future high-mobility channels are forecasted, and it is necessary to have the highest capacitance possible (in other words, thinner interfaces). In these samples, the lowest CET that was obtained with the Ti/Pt gate was of only 2.1 nm, which translates to an EOT of about 1.4 nm. The issue yet to be resolved is a high interface trap density, in the order of $10^{14}eV^{-1}cm^{-2}$ (30), which among other effects, produces a severe flat band voltage displacement with frequency (32). In any case, these devices have no interfacial preparation treatments, so we expect to reduce this interface trap density by passivating the interface by plasma treatments or sulfur passivation (33).

**Summary and conclusions**

In this paper we discussed results on high permittivity dielectrics deposited by a nonstandard method: high pressure sputtering. The dielectrics studied are gadolinium scandates deposited from dielectric targets, and gadolinium oxide deposited from a metallic target and a two-step procedure. With Gd-rich scandate we obtained devices with a high effective permittivity of 21 and no noticeable interface in contact with Si. On the other hand, the devices fabricated with the metallic Gd target present a subnanometric interface that can be further reduced by scavenging with Ti gates. In fact, this scavenging effect has also been demonstrated for the first time with InP substrates.

To summarize, we have tried to answer the question posed on the title. To do that, we have presented results that prove that HPS is a technique that produces reasonably well behaved high-*k* dielectrics, so we can conclude that, indeed, it was definitely worth trying.

**Acknowledgments**

The authors wish to acknowledge CAI de Técnicas Físicas of the Universidad Complutense de Madrid for technical support and CAI de Espectroscopía. The microscopy works have been conducted in the "Laboratorio de Microscopias Avanzadas" at "Instituto de Nanociencia de Aragon - Universidad de Zaragoza". Authors acknowledge the LMA-INA for offering access to their instruments and expertise. This work was funded by the project TEC2010-18051 from the Spanish MINECO, and the FPI program under grant BES-2011-043798.

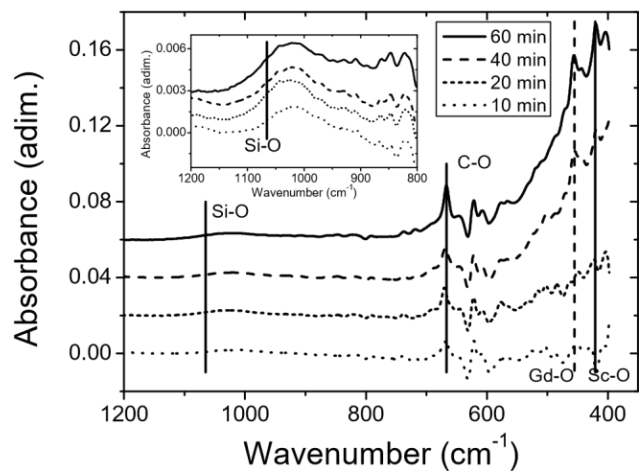

Figure 1: FTIR spectra of $Gd_xSc_{2-x}O_3$ in the 400-1200 cm$^{-1}$ range of films deposited for 10 – 60 min at 200ºC and a pressure of 1.0 mbar of pure Ar. Spectra are corrected with a Si substrate spectrum and they are separated for clarity. Inset: detail of the 800-1200 cm$^{-1}$ range to observe the $SiO_x$ absorption.

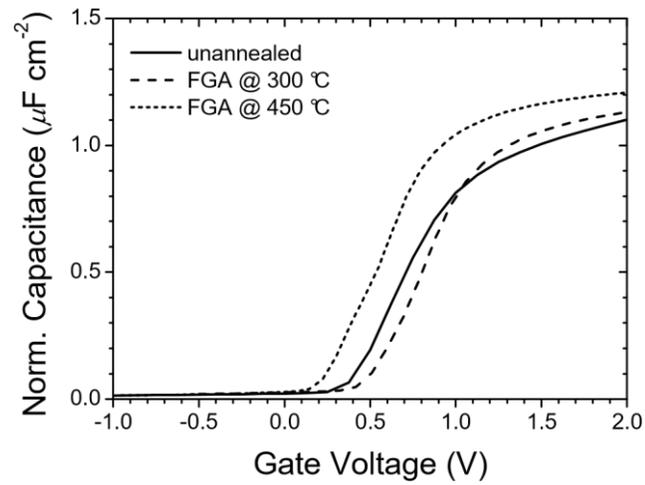

Figure 2: Area normalized gate capacitance at 100kHz as a function of gate voltage of MIS devices with $Gd_xSc_{2-x}O_3$ deposited on Si for 5 min at 200°C and a pressure of 1.0 mbar of pure Ar with pure Pt gate, before annealing (solid lines), and after a 300°C FGA (dashed lines) and 450°C FGA (dotted lines).

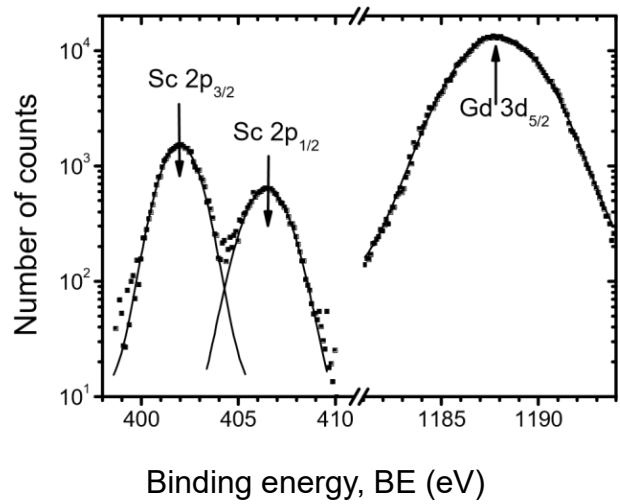

Figure 3: Sc 2p$_{3/2}$, Sc 2p$_{1/2}$ and Gd 3d$_{5/2}$ XPS peaks of a Gd-rich nano-laminate grown by sequential deposition.

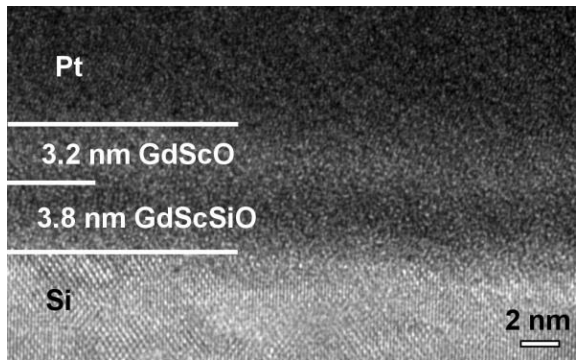

Figure 4: TEM micrograph of the $Gd_{1.8}Sc_{0.2}O_3$ film after the annealing at 450°C in forming gas. The darker interface under the gadolinium scandate suggests the formation of a cuaternary gadolinium-scandium silicate.

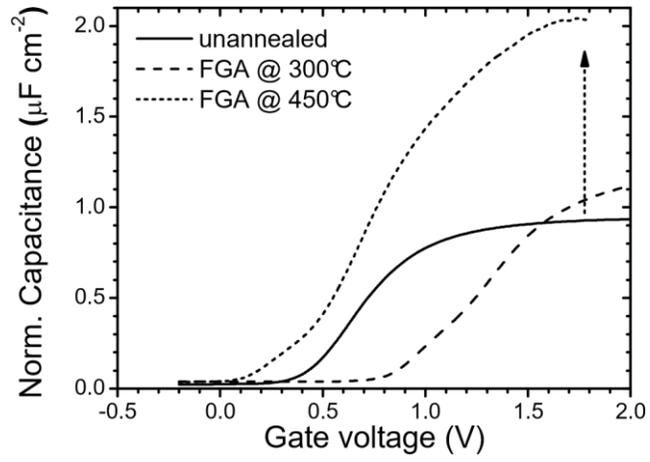

Figure 5: $C_{HF}$–$V_G$ characteristics of representative Pt/Gd$_x$Sc$_{2-x}$O$_3$/Si MIS devices measured at 100 kHz, after device fabrication (solid line), after FGA at 300ºC (dashed line) and after the FGA at 450ºC (dotted line).

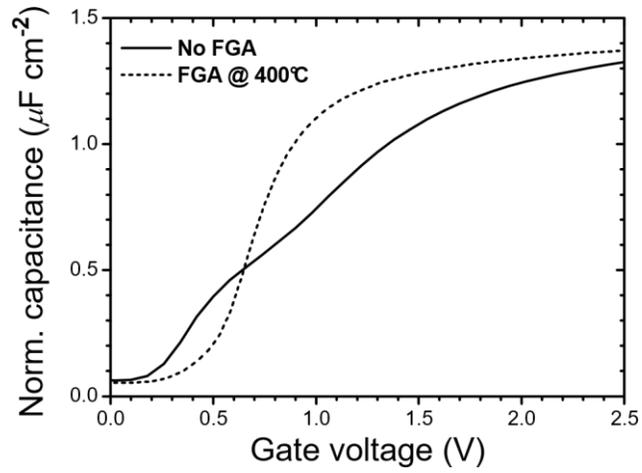

Figure 6: Area normalized gate capacitance at 100 kHz vs gate voltage of Pt/$Gd_2O_3$/Si MIS devices deposited on Si. The $Gd_2O_3$ film was obtained after metal sputtering during 80 s and *in-situ* plasma oxidation during 100 s. The curves correspond to as deposited devices and after FGA at 400ºC.

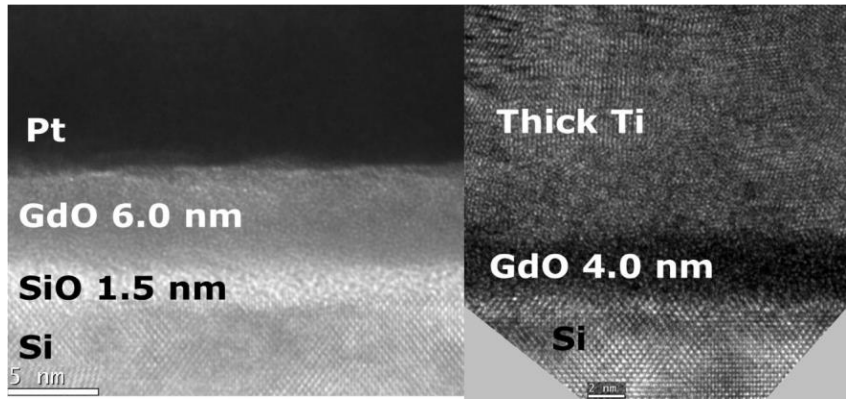

Figure 7: Cross-sectional HRTEM image of a MIS device with FGA at 300ºC and Pt gate electrode (left) or Ti gate electrode (right). $Gd_2O_3$ was obtained after 80 s metal deposition and a 300 s plasma oxidation.

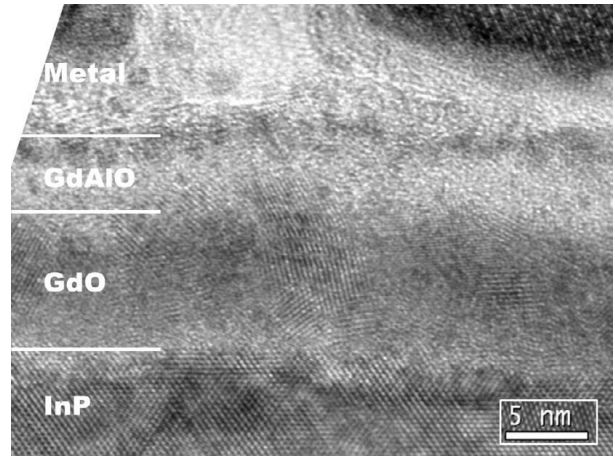

Figure 8: HRTEM image of a MIS device grown on InP with Pt 4 nm/Al 25 nm gate electrode and HPS $Gd_2O_3$ obtained after 120 s metal deposition and a 200 s plasma oxidation.

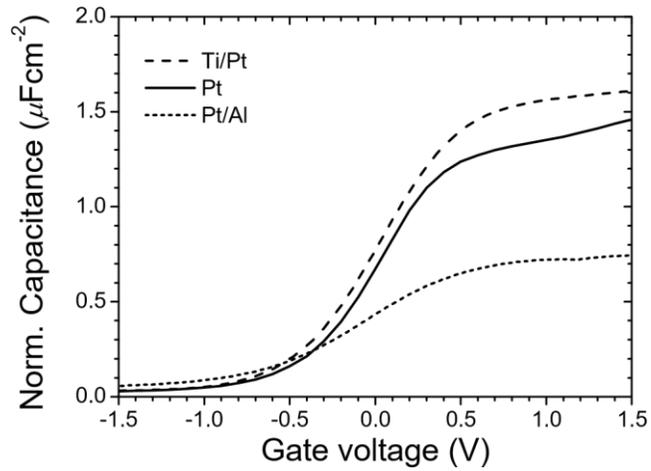

Figure 9: Area normalized gate capacitance measured at 100 kHz as a function of gate voltage of Gd$_2$O$_3$ films deposited by HPS on InP. In all three curves the metallic Gd was deposited during 120 s, while the devices with thin Pt/Al electrodes were oxidized during 200 s and annealed at 400ºC, while the Pt and Ti/Pt devices were oxidized during 100 s and annealed at 325ºC.